\documentclass[aps,physrev,preprint,groupedaddress]{revtex4-2}
\usepackage{amssymb}
\usepackage{amsmath}
\usepackage{amsfonts}
\usepackage[colorlinks,linkcolor=red,urlcolor=blue,citecolor=blue]{hyperref}
\usepackage{graphics}
\usepackage{tikz}  
\usetikzlibrary{arrows,shapes,positioning,shadows,backgrounds,fit}
\usepackage{braket}
\usepackage{tensor}
\usepackage{mathrsfs}

\begin{document}

\title{Vaidya-Reissner-Nordström Extension On the White-hole Region}

\author{Qingyao Zhang}

\affiliation{Mathematical Sciences Institution and Research School of Physics, The Australian National University, Canberra, Australian Capital Territory, 2600, Australia}

\begin{abstract}\noindent
We develop an analytic model that extends classical white hole geometry by incorporating both radiative dynamics and electric charge. Starting from a maximal analytic extension of the Schwarzschild white hole via Kruskal Szekeres coordinates, we introduce a time dependent mass function, representative of outgoing null dust to model evaporation. Building on this foundation, the study then integrates the Reissner-Nordström framework to obtain a dynamic, charged white hole solution in double null coordinates. In the resulting Vaidya Reissner Nordström metric, both the Bondi mass and the associated charge decrease monotonically with retarded time, capturing the interplay of radiation and electromagnetic effects. Detailed analysis of horizon behavior reveals how mass loss and charge shedding modify the causal structure, ensuring that energy conditions are preserved and cosmic censorship is maintained.
\end{abstract}

\maketitle

\section*{Introduction}
\label{intro}
In the framework of general relativity, white holes emerge as the time-reversed analogues of black holes and can be rigorously defined through the maximal analytic extension of classical solutions such as the Schwarzschild metric. Although white holes have traditionally been regarded as mere mathematical  due to their unconventional causal structure, recent theoretical developments have underscored their potential significance in understanding the interplay between gravitation, electromagnetism, and radiative processes.

This paper advances the study of white hole spacetime by constructing a model that unifies the Vaidya metric, which encapsulates radiative dynamics via outgoing null dust, with the Reissner–Nordström solution, which introduces the effects of electric charge in a spherically symmetric setting. By formulating the problem in double-null coordinates and employing retarded time as a natural parameter, we derive a dynamic metric that characterizes a charged, non-rotating white hole whose mass and charge evolve monotonically over time. Such an approach allows us to address the influence of radiation on the spacetime geometry, particularly on the horizon structure and the global causal properties of the system.

Our analysis reveals that the Vaidya–Reissner–Nordström metric not only generalizes the static Reissner–Nordström geometry but also provides an effective framework for modeling the evaporation process of white holes. We generalized Hiscock evaporation black hole model in the white hole situation. The evolution of the Bondi mass and charge under outgoing radiation is shown to preserve the weak energy condition, thereby ensuring consistency with cosmic censorship.

By rigorously exploring these dynamic processes, the present work contributes to a deeper understanding of radiative charged spacetime\cite{Black_to_white} and sets the stage for further investigations into the quantum and astrophysical implications of white hole phenomena. The paper is structured as follows: Section I revisits the Schwarzschild white hole and its maximal extension; Section II introduces the radiative Vaidya metric and its application to white hole dynamics;Section III illustrate the Vaidya white hole evaporation process and dynamic; Section IV develops the extension to include electric charge via the Reissner–Nordström framework; and Section V examines the evaporation dynamics and horizon behavior for the Vaidya-Reissner–Nordström white hole.

\section{Schwarzschild white hole}
We need to start with a basic Schwarzschild white hole, by apply the Schwarzschild solution which with maximal extended. The white hole is the time-reverse edition for the black hole, The metric can be express as:

\begin{equation}
    ds^2=-(1-\frac{2M}{r})dt^2+(1-\frac{2M}{r})^{-1}dr^2+r^2d\Omega^2
\end{equation}
Where, $\Omega^2=d\theta^2+\sin^2\theta d\phi^2$ . 

The Schwarzschild metric have two singularities, $r=0$ and $r=2M$. $r= 2M$ is the coordinate-set singularity, and $r=0$ is the true singularity. To avoid the coordinate singularity, we need to enroll the Kruskal–Szekeres coordinate. First we apply the tortoise coordinate\cite{JBGriffiths2009}, $r^{\star}=r+2M\ln{|\frac {r}{2M}-1|}$ and $dr^{\star}= \frac{dr}{1-\frac{2M}{r}}$, This coordinate stretches the radial coordinate so that the horizon $r=2 M$ been sent to $-\infty$ in $r^{\star}$, and the null coordinate we apply:

\begin{equation}
    u=t-r^{\star}(retarded),v=t+r^{\star}(advanced).
\end{equation}
For the Kruskal-Szekeres coordinate, we enroll:

\begin{equation}
    U=-e^{-\kappa u}, V=e^{\kappa v}.
\end{equation}
Where $\kappa = \frac{1}{4M}$ is the surface gravity. Then by the coordinate transfer, we could get the Schwarzschild metric in Kruskal–Szekeres coordinate as:

\begin{equation}
    ds^2=-\frac{32M^3}{r}e^{-\frac{r}{2M}}dUdV+r^2d\Omega^2,
\end{equation}
where $r$ could be defined implicit by:

\begin{equation}
    UV=(1-\frac{r}{2M})e^{\frac{r}{2M}}
\end{equation}
And we define the timelike coordinate $T$ and spacelike coordinate $X$, when during the situation $r>2M$:

\begin{equation}
    T=2M(V+U), X=2M(V-U),
\end{equation}
when $r<2M$:

\begin{equation}
    T=2M(V-U), X=2M(V+U).
\end{equation}
Then apply to the Eq.(4), the Kruskal-Szekeres coordinate is shown as:

\begin{equation}
    ds^2=\frac{32M^3}{r}e^{-\frac{r}{2M}}(-dT^2+dX^2)+r^2d\Omega^2.
\end{equation}
The coordinate which substitute back, when $r>2M$:
\begin{equation}
\begin{aligned}
& T=4 M \mathrm{e}^{r / 4 M} \sqrt{\frac{r}{2 M}-1} \sinh \left(\frac{t}{4 M}\right), \\
& X=4 M \mathrm{e}^{r / 4 M}\sqrt{\frac{r}{2 M}-1}  \cosh \left(\frac{t}{4 M}\right). \\
\end{aligned}
\end{equation}

When $r<2M$:
\begin{equation}
\begin{aligned}
& T=4 M \mathrm{e}^{r / 4 m}\sqrt{1-\frac{r}{2 M}} \cosh \left(\frac{t}{4 M}\right), \\
& X=4 M \mathrm{e}^{r / 4 M} \sqrt{1-\frac{r}{2 M}} \sinh \left(\frac{t}{4 M}\right).
\end{aligned}
\end{equation}

The transformation will clash when $r=2M$ for curvature singularity in Eq.(9) and Eq.(10).

Then we now consider the maximal analytic extension for the Schwarzschild solution, the Kruskal-Szekeres coordinate could naturally divided by four regions in figure 1.

\begin{figure}
    \centering
    \includegraphics[width=1\linewidth]{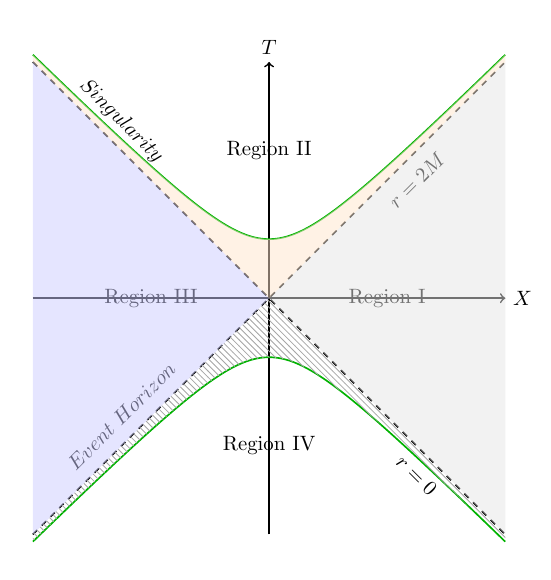}
    \caption{The whole Schwarzschild spacetime in Kruskal–Szekeres coordinate $(T,X)$, the dash line be the event horizon ($r=2M$), two lines which asymptotically to dash lines are the singularity($r=0$), the region IV which filled with slash lines is the white hole region.} The region I($U<0,\ V>0,\ r>2M$) is the exterior region, usually treat as the universe we are in. The region II ($U=0,\ V<0,\ r<2M$) is the black hole region, in this region, all future-directed timelike paths end at the singularity. The region III($U>0,\ V<0,\ r>2M$) is another exterior region, we may interpreted as another universe.
    \label{fig:SchwarzschildKS}
\end{figure}

\begin{figure}
    \centering
    \includegraphics[width=0.8\linewidth]{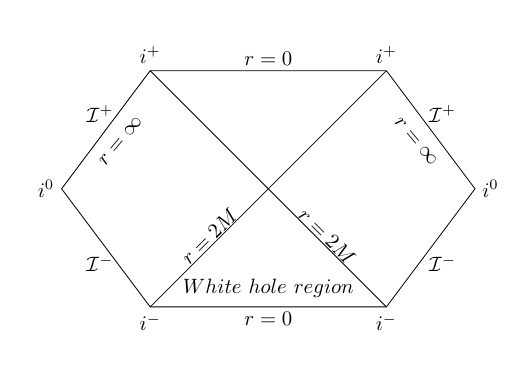}
    \caption{Penrose diagram}
    \label{fig:PDSchWH}
\end{figure}
We remarked the region IV, the white hole interior region, $(r<2M)$ where $U>0$ and $V>0$. This region is the time-reverse of the black hole interior, all the past-directed path originate at the singularity.

The Penrose-Carter diagram of the white hole region for the Schwarzschild spacetime shown as figure 2.
\newpage
\section{Vaidya white hole}
To begin constructing the Vaidya white hole, we start with the Vaidya metric written in the radiation coordinates $(r, \theta, \phi, \omega)$ \cite{Vaidya1951}:
\begin{equation}
    d s^2=2 \epsilon d r d \omega-\left(1-\frac{2 M(\omega)}{r}\right) d \omega^2+r^2 d \Omega^2
\end{equation}

with $\epsilon=1$ for advanced time and $\epsilon=-1$ for retarded time. 

Since we aim to reach the non-static property of the Vaidya white hole, we note that the Kruskal-Szekeres and Eddington-Finkelstein coordinate do not behave well in the time-reversed dynamic region. 

More detailed, although the Kruskal-Szekeres coordinate system is a form of double-null coordinates, it was specifically designed to extend the Schwarzschild metric, and the Eddington-Finkelstein coordinates are less effective in studying the causal structure and dynamic evolution of spacetime in a symmetric null framework. 

Therefore, we apply a general double-null coordinate system. In our study of the white hole, we use retarded time because of the mass is outgoing. And the mass function $M(\omega)$ is monotonically decreasing, $\dot{M}=\frac{\partial M(\omega)}{\partial t}\leq0$\cite{GVS}. The construction in the double-null coordinate as the form\cite{Doublenull}:
\begin{equation}
ds^2=-2f(u,v)dudv+r^2(u,v)d\Omega^2,
\end{equation}

where $d\Omega^2 = d\theta^2 + \sin^2\theta d\phi^2$ is the unit 2-sphere, $r(u,v)$ is the areal radius, and $f(u,v)$ is an undetermined metric function to be solved for later. 

Given $f>0$ to remain the metric signature be $(-,+,+,+)$. The coordinates $u$ and $v$ are chosen to be null, so surfaces of constant $u$ or $v$ are null hypersurfaces.

For an outgoing radiation field, we take $\epsilon=-1$, meaning the null dust flows along the $u$-direction and the mass decreases with increasing retarded time $u$. The stress-energy tensor is that of a single directional null dust:
\begin{equation}
T_{a b}=\rho(u) k_a k_b,
\end{equation}

with $k_a$ a null vector field tangent to outgoing radial null geodesics and we have $k_a \propto \nabla_a u$. We take $k_a=\nabla_a u$, by set the normalization $k_u=1$ and all other $k_v, k_\theta, k_\phi=0$. Hence $T_{u u}=\rho(u)$. 

$T_{ab}$ only has a $uu$-component, consistent with purely outgoing radiation. Physically, $\rho(u)\ge 0$ is related to the mass loss rate. In particular, we assume a mass function $M(u)$ is monotonically decreasing in $u$. The standard Vaidya relation for mass loss:

\begin{equation}
\frac{d M(u)}{d u}=-4 \pi r^2 T_{uu}<0,
\end{equation}

so that $T_{uu} = \frac{-M'(u)}{4\pi r^2}$ (with $M' = dM/du$) is the energy density of the outgoing null flux.\cite{Vaidya1999} All other $T_{ab}$ components vanish, and $M(u)$ will turn out to be the Bondi mass as measured at null infinity. 

Given the above ansatz, we now write down the nontrivial Einstein field equations. Spherical symmetry and the null dust form of $T_{ab}$ yield a system of equations for $f(u,v)$ and $r(u,v)$. In vacuum situation $T_{uu}=0$, we have:
\begin{equation}
r_{u u}=\frac{f_{u}}{f} r_{u}
\end{equation}
where noted $r_{u}=\partial r/\partial u$, $r_{v}=\partial r/\partial v$. The Einstein equations with outgoing null dust $T_{a b}=\rho(u) \nabla_a u \nabla_b u$ given:
$$
G_{u u}=-\frac{2}{f}\left(r_{u u}-\frac{f_u}{f} r_u\right)=8 \pi T_{u u} \Longrightarrow r_{u u}-\frac{f_u}{f} r_u=-4 \pi f T_{u u}(u)
$$

$$
G_{u v}=-\frac{2\left(r_u r_v+r r_{u v}\right)}{f r}+\frac{f}{2}=0 \Longrightarrow 2 \frac{r_u r_v+r r_{u v}}{f}=f\frac{r}{2}.
$$

And we have:
\begin{equation}
2 \frac{r_{u} r_{v}}{f}=\frac{2 M(u, v)}{r}-1,
\end{equation}
where we have defined a local mass function $M(u,v)= \frac{r_{u}^{}r_{v}^{}r}{f}+\frac{r}{2}$. In fact, the above can be recognized as the defining equation for the Misner–Sharp mass in spherical symmetry. Rearranging, one convenient definition is\cite{Hayward}:
\begin{equation}
1-\frac{2 M(u, v)}{r} \equiv-\frac{2 r_{u}^{} r_{v}^{}}{f}
\end{equation}

Using this definition, one can show that $M(u,v)$ is constant along outgoing rays. In the outgoing Vaidya case, one of the Einstein equations implies $\partial_v M = 0$, so that the mass function depends only on $u$. We therefore set $M(u,v)=M(u)$ henceforth. The remaining field equation is the $uu$-component, which relates the change of $M(u)$ to the metric functions. Plugging $M=M(u)$ into the local mass definition and using Eq.(15), one finds the radial equation of motion for outgoing geodesics. It can be written as a first-order PDE for $r(u,v)$:
\begin{equation}
\frac{\partial r}{\partial v}=B(u)\left(1-\frac{2 M(u)}{r}\right).
\end{equation}

Equation (18) is separable and can be formally integrated for general $M(u)$. Treating $u$ as a parameter, we integrate the radial equation as Eq.(19):

\begin{equation}
\int \frac{d r}{1-\frac{2 M(u)}{r}}=\int B(u) d v
\end{equation}

Here $B(u)$ is an arbitrary function of $u$ arising from integration. 

Equation (18) is precisely the single first-order nonlinear PDE that the Einstein equations reduce to for the Vaidya metric. It governs how the areal radius $r(u,v)$ evolves as one moves along increasing $v$, for each fixed $u$. $B(u)$ can be chosen positive so that $v$ increases toward the future for each fixed $u$. Performing the $r$-integral in Eq.(19) one obtains:

\begin{equation}
r-2 M(u)+2 M(u) \ln |r-2 M(u)|=B(u) v+C(u),
\end{equation}

Here $C(u)$ is an arbitrary integration function of $u$, where the additive constant $-2 M(u)$ may be absorbed into the arbitrary function $C(u)$. A suitable choice of $C(u)$ will be made to ensure the coordinates are regular across any horizon. 

Now, with $r(u,v)$, we can construct the metric function $f(u,v)$. One convenient way to obtain $f$ is to use the relation $g^{uv} = -1/f$ together with the definition of $M(u)$. Equivalently, we can differentiate the implicit solution above with respect to $v$. Differentiating yields back the radial PDE (19), $r_{v}^{} = B(u)\big(1 - 2M/r\big)$. Thus, solving for $f = -g_{uv}$:
\begin{equation}
f(u, v)=\frac{r_v}{B(u)},
\end{equation}

so that,

\begin{equation}
f(u, v)=1-\frac{2 M(u)}{r(u,v)}.
\end{equation}

This is the desired metric coefficient in double-null coordinates, valid for an arbitrary mass function $M(u)$. $B(u)$ gauges the normalization of $v$ along each outgoing null cone with fixed $u$. $C(u)$ gauges the origin of $v$ on each cone. Neither of them changes the scaling of the retarded coordinate $u$.

One may now exploit the coordinate freedom encoded in the functions $B(u)$ and $C(u)$. First, fix the overall scaling of the ingoing coordinate $v$ by choosing $B(u)=1$. Which makes the retarded coordinate $u$ coincide with the usual Bondi retarded time at future null infinity. Second, use the additive freedom $v \rightarrow v+\sigma(u)$ to select $C(u)$ so that the metric remains analytic at the outgoing apparent (or event) horizon $r=2 M(u)$; in the static Schwarzschild case this gives the familiar Kruskal choice $C(u)=-2 M$.

With those choices the implicit solution of the radial equation is:
$$
r+2 M(u) \ln |r-2 M(u)|=v,
$$

and the metric coefficients are:

\begin{equation}
\begin{aligned}
    &d s^2=-2\left(1-\frac{2 M(u)}{r}\right) d u d v+r^2 d \Omega^2,\\
    &f(u, v)=1-\frac{2 M(u)}{r(u, v)}.
\end{aligned}
\end{equation}

In the static limit $M(u)=M$ these reduce to the advanced Eddington-Finkelstein form; with a different $C(u)$ one recovers the Kruskal-Szekeres extension. For a dynamical $M(u)$, a suitably chosen $C(u)$ likewise ensures a smooth extension across any future (white-hole) horizon that forms.

Together with the implicit relation for $r(u,v)$ in Eq.(20), this describes a general radiating white-hole spacetime. The outgoing Vaidya white hole metric is often presented in retarded null coordinates $(u,R)$. To recover that form, one can use the freedom in $v$ to set $C(u)$ such that $v$ coincides with the ingoing null coordinate of flat infinity. Then $u$ is our retarded time, and $R=r$ is the areal radius. The metric becomes:
\begin{equation}
d s^2=-\left(1-\frac{2 M(u)}{R}\right) d u^2  -2 d u d R+R^2 d \Omega^2,
\end{equation}

which indeed is the standard outgoing Vaidya metric (with $M(u)$ decreasing) describing a radiating white hole or evaporating black hole. In our double-null coordinates, the same physics is encoded but the coordinates remain regular at the outgoing horizon by construction. The function $C(u)$ has been chosen so that $(u,v)$ extend through $r=2M(u)$, unlike the simpler $(u,R)$ coordinates which break down at the horizon.

In summary, we started from Einstein’s equations with a null dust stress-energy and arrived at a single first-order PDE for $r(u,v)$. Its integration introduced an arbitrary function of $u$, which we denoted $C(u)$, that must be specified by a regularity condition. The general solution for arbitrary $M(u)$ is given implicitly by the relations above. For the white-hole case, $M(u)$ is a decreasing function (outgoing positive energy flux), and one picks the integration constants to produce an outgoing radiation solution regular on the outgoing horizon. Our construction generalizes the Schwarzschild-Kruskal coordinates to the dynamic case of a changing mass. The result is a double-null Vaidya metric\cite{Doublenull}.

\section{Vaidya white hole evaporation}

\subsection{Basic Structure}

To analyze the global structure, we write a double-null general spherically symmetric metric which developed from the last chapter as:
\begin{equation}
d s^2=-e^{2 \sigma(u, v)} d u d v+r^2(u, v) d \Omega^2,
\end{equation}

where $\sigma(u, v)$ is a conformal factor which arbitrary function for $u$ and $v$. $u$ and $v$ are independent null coordinates. The Vaidya solution can transfer in this form, but it is often simpler to use one of the Eddington–Finkelstein forms above with an explicit mass function. In these coordinates, Einstein’s equations can be solved given a specified $M(u)$ or $M(v)$, and the stress-energy will correspond to a null dust flow. Below we focus on the outgoing Vaidya metric for an evaporating white hole with $M = M(u)$ decreasing. 

The function $M(u)$ appears such that in the static limit ($M=\text{const}$) we recover Schwarzschild mass. Geometrically, one can define a local mass function $m(u,r)$ via the equation $1-\frac{2m(u,r)}{r} = g^{\mu\nu}(\partial_\mu r)(\partial_\nu r)$. In Vaidya, $m(u,r)$ is simply $M(u)$ (independent of $r$ in the exterior), meaning the spacetime’s effective mass at a given retarded time is $M(u)$. Einstein’s field equations $G_{ab}=8\pi T_{ab}$ then relate the $u$-derivative of $M$ to the stress-energy of the null dust. In fact, the Vaidya metric with $M(u)$ is an exact solution with Ricci tensor and stress-energy concentrated along null directions. For the outgoing metric one finds the only nonzero Ricci component is $R_{uu} = -\frac{2M'(u)}{r^2}$, and all other components and the Ricci scalar vanish except at $r=0$. The Einstein equations then give the stress-energy tensor $T_{ab}$ as a pure null dust:

\begin{equation}
T_{a b}=-\frac{M^{\prime}(u)}{4 \pi r^2} l_a l_b,
\end{equation}

where $l_a$ is a null covector aligned with $du$ (one can take $l_a dx^a = -du$, so $l_a = -\partial_a u$). Physically, $l^a$ points along the outgoing null rays with radial lightlike direction. The energy density measured by an observer on this null flow is:
\begin{equation}
    \rho = -\frac{M'(u)}{4\pi r^2}.
\end{equation}
For the stress-energy to represent positive energy flux leaving the object, we take

$$
\rho=-\frac{M^{\prime}(u)}{4 \pi r^2}>0 \quad \Longrightarrow \quad M^{\prime}(u)<0 .
$$

The null-energy condition

$$
T_{a b} k^a k^b \geq 0
$$

with $T_{a b}=-\frac{M^{\prime}(u)}{4 \pi r^2} l_a l_b$ therefore requires
$M^{\prime}(u) \leq 0$; for the evaporating case $M^{\prime}(u)<0$ and the NEC is satisfied.
In the ingoing Vaidya metric one instead has

$$
T_{a b}=\frac{M^{\prime}(v)}{4 \pi r^2} n_a n_b
$$

Here the NEC implies $M^{\prime}(v) \geq 0$, describing accretion or collapse. Forcing the mass to decrease in this chart $\left(M^{\prime}(v)<0\right)$ would indeed require negative energy density and violate the classical energy conditions an idealized stand-in for quantum effects such as Hawking radiation, much as the Casimir vacuum violates point wise energy conditions. And the violation for the energy condition is not impossible, experimentally shown in Casimir effect by \cite{Casimir1}\cite{M.V}\cite{Birrell.D.1982}\cite{PLUNIEN198687}\cite{Visser1995cc}.

\subsection{Mass Loss and Outgoing Energy Flux with Hiscock Model}
In the outgoing Vaidya picture, $M(u)$ encapsulates the evaporation law. For example, one might prescribe a linear decrease $M(u)=M_0 - \alpha u$ during some interval or an asymptotic decay $M(u)\sim M_0 - \beta\ln(1+u)$, etc. The rate $|M'(u)|$ is related to the power carried by the radiation. In particular, far from the hole, the luminosity measured at infinity is $-\frac{dM}{du}$ (since $M'(u)<0$). This matches the idea that the mass-energy radiated to infinity reduces the Bondi mass at future null infinity. The outgoing energy flux density is $T_{u u}=-M^{\prime}(u) /\left(4 \pi r^2\right)$, integrating $4 \pi r^2 T_{u u}$ over a sphere at large $r$ reproduces the total luminosity $L(u)=-M^{\prime}(u)$, precisely the Bondi mass-loss rate.\cite{Bondi2003}

W. A. Hiscock (1981) \cite{HiscockI} constructed explicit piece-wise Vaidya space-times to model an evaporating black hole. The geometry is glued together from three null regions.

Region I: an initial Schwarzschild space-time of mass $M_0$.

Region II: an ingoing Vaidya patch with a monotonically decreasing mass function $M(v)$ (advanced time), representing a stream of negative-energy null dust that encodes the Hawking back-reaction.

Region III: flat Minkowski space reached when $M(v) \rightarrow 0$.

This shown as figure 4.
\begin{figure}
    \centering
    \includegraphics[width=0.5\linewidth]{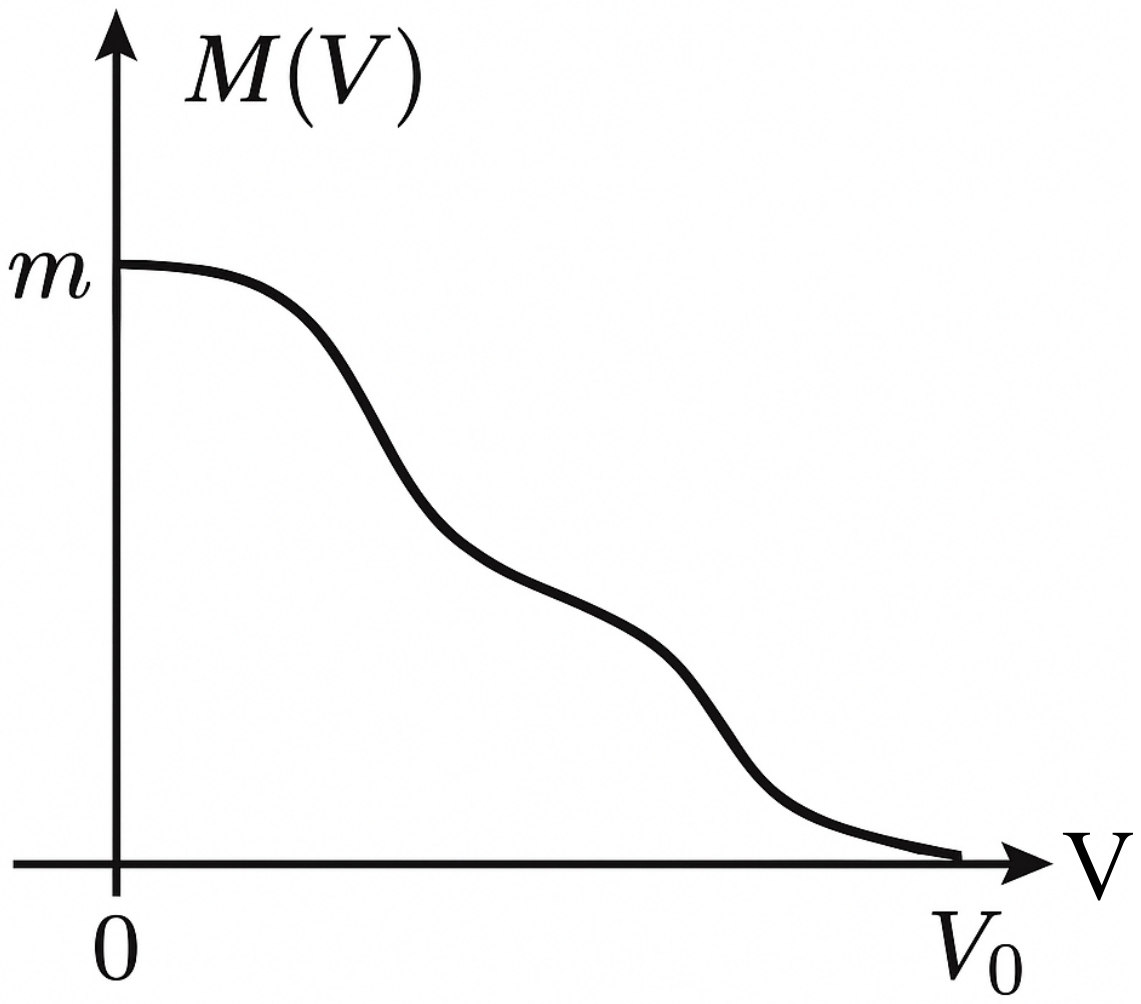}
    \caption{Choice for $M(v)$ for the Hiscock model\cite{HiscockI}}
    \label{fig:enter-label}
\end{figure}
Because $M(v)$ vanishes at a finite advanced time $v_{\text {end }}$, the event horizon exists only for a finite interval and terminates at a point, beyond which a Cauchy horizon forms. Hiscock showed that for any evaporation law that brings the mass to zero in finite time-even a smooth linear decay-the renormalised stress tensor of a quantum test field diverges as

\begin{equation}
    \left\langle T_{a b}\right\rangle \propto\left(v_{\text {end }}-v\right)^{-2}
\end{equation}

on that Cauchy horizon, signaling an infinite burst of outgoing particles. The lesson is that a self-consistent semiclassical evaporation should have

\begin{equation}
    M(v) \longrightarrow 0 \text { only as } v \rightarrow+\infty,
\end{equation}

so that the horizon "evaporates away" gradually without generating unbounded energy densities.

Nonetheless, as a classical model, one can choose either case: complete evaporation in finite time (with $M(u_{\text{end}})=0$ for some finite $u_{\text{end}}$) or evaporation in infinite time ($M(u)\to0$ as $u\to\infty$). We will discuss how each affects the horizon structure and curvature. In any case, the local mechanism is that outgoing null radiation (null dust) carries away mass-energy, reflected in $M'(u)<0$ in the Einstein equations. Mathematically, the decrease of $M(u)$ directly enters the metric and Christoffel symbols, affecting the focusing of geodesics. Because $T_{ab}$ has only a null-null component, the dust has zero rest mass and moves at $c$. The local conservation $\nabla^a T_{ab}=0$ is satisfied automatically for the chosen $M(u)$ (except at $r=0$ where a source or sink would formally reside). 

One can check that the mass loss rate equals the flux through a sphere: 
\begin{equation}
    -dM/du = \int T_{0r}r^2 d\Omega,
\end{equation}
at infinity. Thus $M(u)$ decreases exactly in accordance with the radiated energy. From a double-null formalism perspective, one often looks at the null expansions and the Raychaudhuri equation\cite{Raychaudhuri}. The outgoing null congruence in the exterior has expansion $\theta_{(\ell)}>0$ (since the area of light fronts increases), while the ingoing congruence ($n^a$) in a white-hole spacetime can have positive or negative expansion depending on region. In fact, for the outgoing Vaidya metric one finds outgoing null expansion is:
\begin{equation}
    \theta_{(\ell)} = \frac{2}{r},
\end{equation}
where $\theta_{(\ell)}$  is the expansion scalar for the outgoing null congruence and it always positive outside the singularity, the ingoing null expansion is:
\begin{equation}
    \theta_{(n)} = \frac{-r + 2M(u)}{r^2},
\end{equation}
where, $\theta_{(n)}$ is the expansion scalar for the ingoing null congruence.

At large $r$, $\theta_{(n)}\approx -1/r < 0$ which ingoing rays converge as in flat space. But if we approach the radius $r = 2M(u)$, the ingoing expansion $\theta_{(n)}$ approaches zero and changes sign. Inside the white hole, one actually has $\theta_{(n)}>0$ (an anti-trapped region where even ingoing light is forced outward). The surface $r = 2M(u)$ thus acts as a marginal surface. In fact, $r=2M(u)$ is a dynamical apparent horizon for the spacetime: it is the boundary between normal space and the anti-trapped region. For the ingoing Vaidya black hole case, $r=2M(v)$ is a marginally trapped surface. In either case, the Vaidya horizon satisfies conditions like $\theta_{(\ell)}\theta_{(n)}=0$ on it. We identify $r_h(u) = 2M(u)$ as the apparent horizon radius at a given retarded time $u$. Differentiating this:
\begin{equation}
\frac{d r_h}{d u}=2 \frac{d M(u)}{d u}=2 M^{\prime}(u)
\end{equation}

Since $M'(u)<0$ for evaporation, we have $dr_h/du < 0$, the horizon radius shrinks over time. This has profound implications for the nature of the horizon, it is no longer a stationary lightlike surface as in Schwarzschild, but instead moves inward.
\subsection{Horizon Dynamics and Spacetime Structure during Evaporation}

In a dynamical space-time the event horizon is defined globally, whereas an apparent (or trapping) horizon is defined locally.
For the outgoing-Vaidya white hole the natural apparent/anti-trapping horizon is the surface
\begin{equation}
    r=2 M(u)
\end{equation}
which is a timelike past outer trapping horizon.
If the evaporation leaves no remnant, the event horizon initially coincides with this surface but terminates when the mass function reaches zero. Concretely, if $M(u)$ vanishes at some finite retarded time $u_{\text {end }}$, then for $u>u_{\text {end }}$ the space-time is flat Minkowski and contains no trapped or anti-trapped surfaces; the event horizon therefore ends at that point.

If instead $M(u)$ tends to zero only asymptotically as $u \rightarrow \infty$, the apparent and event horizons shrink continuously to zero radius in the infinite future.

In Schwarzschild, the event horizon is a null hypersurface. But with a changing mass, the horizon can tilt off the lightlike direction. Analytical studies show that if the mass is evaporating, the horizon tends to be timelike or spacelike rather than null. For a slowly changing mass, one finds the horizon world tube is approximately null but with a slight timelike tilt for an evaporating black hole. In contrast, for a rapidly shrinking white hole, the horizon can become spacelike.

In summary, for evaporating black hole, horizon is found to be timelike in the Vaidya solution (to first order in $M'$). This means an outside observer could actually cross the retreating horizon if they move fast enough outward, since the horizon is moving inward. Intuitively, the horizon is contracting, so outgoing light just outside can outrun it. A timelike horizon also implies it has a surface area that decreases with time, when $dA<0$ we say it as timelike membrane. This is a violation of the classical area theorem, but it is possible here because the null energy condition and non-null horizon is violated by Hawking-like flux.

For shrinking white hole, horizon is spacelike in analyses of dynamic solutions. In the time-forward picture, it means the horizon’s boundary moves at faster-than-light rate inward. In essence, the white hole’s anti-trapping horizon vanishes so quickly that it lies within the backward light cone of its boundary. Physically, once the white hole has emitted most of its mass, the remaining horizon collapses “faster than light” in the diagram, so no world line can remain on the horizon, as it is a fleeting, space-like boundary. This corresponds to a rapid evaporation or an endpoint of the white hole. By contrast, an expanding white hole, which would have increasing $M$ if such were considered, yields a timelike horizon, analogous to an evaporating Black hole’s horizon being timelike.

These properties were confirmed in recent studies of Kerr–Vaidya spacetime\cite{Kerr2025}, which reduce to spherical symmetry in the limit $a\to0$, as $a$ defined as $a=J/M$ is the angular momentum per unit, as non-rotation circumstance, $a=0$. There, it was found that a shrinking white hole’s horizon is spacelike, whereas a slowly expanding one is timelike. Similar, a slowly evaporating black hole’s apparent horizon is timelike. All these dynamic horizons differ from the static case, which the classical event horizon will wrinkles and tilts due to the mass flux.

\subsection{Global structure}

The spacetime of an evaporating white hole can be visualized as follows. Start with a past singularity (much like the white hole singularity in the past of an eternal Schwarzschild). Emanating from this $r=0$ singularity is the white hole event horizon (a past horizon) which initially might be near $r=2M_{\text{initial}}$. As time progresses, the horizon moves outward for a while (if the white hole initially increases to $M_{\text{initial}}$) and then begins to shrink as $M$ decreases. Outside the horizon, there is an outflow of radiation to infinity. If the evaporation is complete, at some retarded time $u=u_+$ the mass approaches zero. At this point, the horizon meets the center $r=0$. In a Penrose diagram, the horizon might terminate at a point on future null infinity or at future timelike infinity depending on how $M$ vanishes. In any case, after evaporation, the spacetime becomes flat or at least free of any horizon or singularity – effectively connecting to a region isometric to a portion of Minkowski space. No future singularity remains, since the original singularity was in the past. In this sense, the white hole has completely disappeared, leaving behind empty space. 

However, an important subtlety is what happens to the region that was formerly inside the horizon. During the white hole’s active phase, the interior was a region of no-return (for past-directed rays) bounded by the horizon. Once the horizon is gone, that interior region becomes accessible. The termination of the horizon can thus create a kind of Cauchy horizon: beyond that point, the evolution of spacetime is no longer determined by initial data on a nice spacelike slice (because part of the domain was previously causally disconnected). In classical models like Hiscock’s, the point where $M=0$ and $r_h=0$ (horizon meeting the axis) is a naked singularity or at least a singular boundary of the spacetime. Indeed, Hiscock found that a zero-mass naked singularity at the end of evaporation is not necessarily benign – it can be accompanied by divergent particle creation. Other analyses have shown that if the evaporation ends abruptly (finite $u$ with nonzero $M'(u_{\text{end}})$), a null curvature singularity may form along the last ray of evaporation. If instead $M(u)$ goes to 0 asymptotically as $u\to\infty$, the spacetime can be extended indefinitely, but even then a sort of asymptotic null singularity in curvature can arise. In one study\cite{GVS}, it was proven that in a complete evaporation in infinite time, there is a lightlike Weyl curvature singularity that extends to future timelike infinity. This is essentially a burst of gravitational tidal force riding out on the final wave of radiation. 

To summarize the global structure, one begins with a past spacelike singularity which also the white hole’s origin. From it emanates a past event horizon. For $u$ in some range $(-\infty, u_-)$, the mass might be roughly constant as $M=M_-$, this would be an initial white hole state. Then for $u_- < u < u_+$, the mass $M(u)$ decreases. During this period, spacetime is Vaidya with an outgoing flux. Incoming light rays near the white hole horizon satisfy an equation of motion that can be solved to track the horizon generators. As $u\to u_+$ which the end of evaporation, $r_h(u)\to 0$. If $M_+=M(u_+)=0$, the horizon closes off at the center and there is no future event horizon, all worldlines after $u_+$ can eventually escape to infinity. The portion of spacetime after $u_+$ is essentially flat (if completely evaporated). The Penrose diagram would resemble Minkowski space on the future, joined to a radiating Vaidya region in the middle, and a triangle representing the white hole interior bounded by the past horizon on the left. All timelike observers survive to infinity; there is no black hole region. The apparent horizon $r=2M(u)$ is a line that starts at $r=2M_-$ at past infinity and curves inward to meet $r=0$ at $u_+$. All trapped/anti-trapped surfaces disappear after that point. If the evaporation is only partial (i.e. $M(u)\to M_+ >0$ as $u\to\infty$), then the white hole eventually stabilizes into a black hole of mass $M_+$. In that case, the past horizon merges into a future horizon of a remaining Schwarzschild black hole. But with complete evaporation with $M_+=0$, the past event horizon vanishes in finite time and no new horizon forms. All that remains is outgoing radiation in an otherwise flat spacetime. 

\section{Vaidya-Reissner-Nordström white hole}

We assume a spherically symmetric double-null metric of the form:

\begin{equation}
d s^2=-2 f(u, v) d u d v+r^2(u, v) d \Omega^2
\end{equation}
with $u$ a retarded null coordinate and $v$ an advanced null coordinate. The functions $M(u)$ and $q(u)$ represent the mass and electric charge inside radius $r$ at retarded time $u$, which will generally decrease as the white hole emits null radiation and discharges. We model the stress-energy as a combination of an outgoing null dust and an electromagnetic (EM) field. The null dust is described by an energy–momentum tensor:
\begin{equation}
    T_{\mu\nu}^{\rm (rad)} = \Psi(u) l_\mu l_\nu
\end{equation}
where, $l^\mu$ a radial null vector along $du$ direction. And the electromagnetic field by the Maxwell tensor $F_{\mu\nu}$ of a purely radial electric field. In spherical symmetry the only nonzero Maxwell components can be taken to be $F_{uv} = F_{vu}$ such that for a static charge $q$ one recovers $F_{tr}=q/r^2$. Maxwell’s equations $\nabla^\nu F_{\mu\nu}=J_\mu$ imply $\partial_u q(u) \neq 0$ requires a radial current $J_\mu$ flowing outward\cite{Chirenti_2012}. In the absence of external currents ($J_\mu=0$) the charge would be conserved ($q(u)=\text{const}$); here we allow $q(u)$ to vary by including an appropriate outgoing current $J_u(u)$ in the retarded direction.

We work in units $8 \pi G=c=1$ and assume $r(u, v)$ and $f(u, v)$ are smooth. With $g_{u v}=-f$, the non-zero Einstein equations along the null directions are\cite{Gundlach_1997}

$$
G_{u u}=T_{u u}, \quad G_{v v}=T_{v v} .
$$

The $uu$ component be:

\begin{equation}
r_{u u}-\frac{f_u}{f} r_u=-\frac{f}{2 r} T_{u u} .
\end{equation}

For our matter content the only $u u$-stress comes from outgoing null dust,

\begin{equation}
    T_{u u}=\Psi(u)\left(l_u\right)^2=\Psi(u), \quad l_\mu=-\partial_\mu u
\end{equation}
Dividing Eq.(37) by $f$ gives the differential form:
\begin{equation}
\partial_u\left(\frac{r_u}{f}\right)=-\frac{\Psi(u)}{2 r(u, v)} .
\end{equation}

Integrate Eq.(39) from a reference retarded time $u_0$ :

\begin{equation}
    \frac{r_u}{f}=A(v)-\int_{u_0}^u \frac{\Psi(\bar{u})}{2 r(\bar{u}, v)} d \bar{u}
\end{equation}

where $A(v)$ is an arbitrary function of the advanced coordinate, fixed by the remaining gauge freedom or initial data.

Multiplying by $f(u, v)$ yields

\begin{equation}
    r_u(u, v)=f(u, v)\left[A(v)-\int_{u_0}^u \frac{\Psi(\bar{u})}{2 r(\bar{u}, v)} d \bar{u}\right]
\end{equation}
In vacuum Eq.(40) reduces to $r_u / f=A(v)$, reproducing the familiar constraint for double null Schwarzschild.

If the radiation flux is expressed in terms of the mass loss, we expect $T_{uu} = \Psi(u)\sim -\partial_u m(u)/ (4\pi r^2)$ by assume $f=1$ In fact, integrating $G_{uu}$ reveals that $m(u)$ is an integration function related to the energy flux. In the pure radiation case ($q=0$), one indeed finds $r_{u} = A(v)f$ (no explicit $u$-dependence aside from $f$) consistent with Vaidya metric in outgoing coordinates.

For $vv$-component,
\begin{equation}
r_{v v}-\frac{f_v}{f} r_v=-\frac{f}{2 r} T_{v v} .
\end{equation}
In our scenario with only outgoing radiation, we have no ingoing flux ($T_{vv}=0$). This integrates similarly to give $\frac{r_{v}^{}}{f}=B(u)$.

One can fix the scaling of $v$ such that $B(u)=1$. Physically, in the absence of ingoing matter, $r_{v}^{}$ can be adjusted by reparameterizing $v$. Below we will often choose this gauge so that $r_{v}=f$ for simplicity.

For $uv$-component, $G_{uv}=T_{uv}$. In our case $T_{uv}$ will come only from the electromagnetic field, since null dust has $T_{uv}=0$ for a purely outgoing flow. The electromagnetic energy-momentum for a radial $E$-field contributes an isotropic pressure. Specifically, one finds $T_{\theta}^{\theta}=T_{\phi}^{\phi}=-T_{u}^{u}=-T_{v}^{v}= \frac{q(u)^2}{8\pi r^4}$ for the Coulomb field. As a result, $G_{uv}=0$ in our setup. Thus the $uv$-equation is a constraint relating $f$ and $r$. Evaluating $G_{uv}=0$ for the metric above yields:
\begin{equation}
    \frac{\partial^2}{\partial u \partial v}(\ln f(u, v))+ \frac{2 r_{uv}^{}}{r} - \frac{2q(u)^2f(u,v)}{r^4} = 0.
\end{equation}

This equation ensures consistency of the double-null coordinates. It is satisfied automatically once the other components and Maxwell’s equations hold. In particular, in vacuum it reduces to:
\begin{equation}
    \frac{\partial^2}{\partial u \partial v}(\ln f(u, v)) + \frac{2r_{uv}^{}}{r}=0,
\end{equation}

which is identically satisfied if $r_{u}^{'}$ and $r_{v}^{'}$ obey the above first integrals and $f$ is solved from the $\theta\theta$-equation below. We note that Eq.(40) is the only mixed second-order equation; accordingly, in the pure Vaidya case ($q=0$) the Einstein equations reduce to a single first-order PDE for $r(u,v)$ (after using the first integrals). With charge present, the structure is similar but includes the $q^2$ term as shown.

For $\theta\theta$-component, $G_{\theta}^{\theta}=T_{\theta}^{\theta}$. This is the crucial equation that relates the cross-derivative $r_{uv}^{'}$ to the metric function $f(u,v)$. Computing $G_{\theta}^{\theta}$ for the metric yields:
\begin{equation}
r_{u v}^{'}+\frac{r_{u}^{'} r_{v}^{'}}{r}-\frac{f(u, v)}{r}+\frac{q(u)^2 f(u, v)}{r^3}=0 .
\end{equation}

This is a key equation governing the geometry. The terms $-f/r + (q^2 f/r^3)$ on the left come from the Einstein curvature. And $-f/r$ is the usual Schwarzschild term, $q^2 f/r^3$ arises from the electromagnetic field’s contribution $T_{\theta}^{\theta}$. Equation (41) can be rearranged to solve for $f(u,v)$ in terms of $r(u,v)$ and its derivatives. Moving the $f$ terms to the right side gives:
\begin{equation}
r_{u v}^{'}+\frac{r_{u}^{'} r_{v}^{'}}{r}=\frac{f(u, v)}{r}\left(1-\frac{q(u)^2}{r^2}\right) .
\end{equation}
Solving for $f(u,v)$, we obtain an expression for $f$:
\begin{equation}
f(u, v)=\frac{r r_{u v}^{'}+r_{u}^{'} r_{v}^{'}}{1-\frac{q(u)^2}{r^2}}
\end{equation}

This formula is completely general for our ansatz, it determines the metric function $f$ once $r(u,v)$ is known. Equation (42) is equivalent to the standard “mass function” relation in spherically symmetric spacetimes. Indeed, one can identify the local mass function $M(u,v)$ by writing $1 - \frac{2M(u,v)}{r} + \frac{q(u)^2}{r^2} = g^{\mu\nu}\partial_\mu r\partial_\nu r$. In double-null coordinates $g^{uv}=-1/f$ and $g^{uu}=g^{vv}=0$, so this becomes $1 - \frac{2M}{r} + \frac{q^2}{r^2} = -\frac{2r_{u}^{'}r_{v}^{'}}{f}$. Using Eq.(42) to eliminate $f$, one finds $2M/r = 1 + \frac{2r_{u}^{'}r_{v}^{'}}{f} + \frac{q^2}{r^2}$, which indeed is consistent with the usual definition of the Misner–Sharp mass in charged spacetime. In particular, one can show $\partial_v M=0$, so that $M=M(u)$ is a function of $u$ alone, this $M(u)$ is precisely the Vaidya–Reissner–Nordström mass function $m(u)$.\cite{Gundlach_1997}

The strategy is to use the simpler $uu$ and $vv$ equations to integrate for first derivatives of $r$, and then use Eq.(41) to identify the integration functions in terms of $m(u)$ and $q(u)$. As noted, from $G_{uu}$ we get $r_{u}^{'}=A(v)f$ and from $G_{vv}$ we get $r_{v}^{'}=B(u)f$. Plugging these into Eq.(41) yields a consistency condition: differentiating $r_{u}^{'}=A(v)f$ with respect to $v$ and using $r_{uv}^{'}$ from Eq.(41) shows that $\partial_v\big(rr_{u}^{'}\big) = f\big(1 - \frac{q^2}{r^2}\big)$. In other words,
\begin{equation}
\frac{\partial}{\partial v}\left(r r_{u}^{'}\right)=\frac{f(u, v)}{r(u, v)}\left(r^2(u, v)-q(u)^2\right)
\end{equation}

If no ingoing radiation is present, one can choose the gauge $B(u)=1$ so that $r_{v}^{'}=f$. Then the above equation simplifies to $\partial_v\big(r,r_{u}^{'}\big)=\partial_v\big(r,A(v)f\big)=f(1-\frac{q^2}{r^2})$. In regions away from horizons ($f\neq0$), this implies $\partial_v\big(r,r_{u}^{'}\big)=0$, so $r,r_{u}^{'}$ is actually independent of $v$. This quantity can only be a function of $u$: we identify it (up to a constant factor) with the mass function. In fact, one finds:
\begin{equation}
r r_{u}^{'}(u, v)=1-\frac{2 m(u)}{r(u, v)}+\frac{q(u)^2}{r(u, v)^2},
\end{equation}

which is consistent with our definition above. Solving for $m(u)$ gives: $2m(u)/r = 1 + rr_{u}^{'} - q(u)^2/r^2$. Evaluated at large $r$ where $rr_{u}^{'}\to 0$ for an asymptotically flat radiating solution, this yields $2m(u)/r \approx 1 - q(u)^2/r^2$, so $m(u)$ is exactly the Bondi mass of the system at retarded time $u$. Differentiating with respect to $u$, we get $\dot m(u) = -\frac{1}{2}\dot q(u)^2 (1/r)$ plus higher order terms, at infinity this reduces to $\dot m(u)\approx -\frac{\dot q(u)^2}{2r}$, which suggests that a changing charge contributes to the energy flux. Indeed, the energy conservation condition ensures that the loss of mass is related to both the outgoing null dust energy and the work done to carry away charge. In the limit of no charge, one finds $\dot m(u) = -\Psi(u)$, consistent with the Vaidya solution where $\Psi(u)$ is the luminosity (energy flux) of the radiation. In the general case, one can similarly relate $\dot m(u)$ to $\Psi(u)$ and $\dot q(u)$ by expanding the field equations, effectively $\dot m(u) = -\Psi(u) - \frac{q(u)\dot q(u)}{r}$, the second term is small far from the hole, but indicates that if charge is being lost, part of the mass loss is due to the EM field’s energy. The weak energy condition requires $\Psi(u)\ge0$ and also restricts how $m(u)$ and $q(u)$ vary. In particular, one finds $m(u)$ must be a non-increasing function (so a white hole’s mass decreases with retarded time) and if both $m$ and $q$ vary, they cannot do so arbitrarily. $\dot m(u)$ and $\dot q(u)$ must satisfy an inequality to keep $T_{\mu\nu}k^\mu k^\nu\ge0$ for all null $k^\mu$. Physically, if charge is radiated away, it must be accompanied by energy loss. In the simplest case where only mass varies (charged radiation absent), $\dot m(u)\le0$ monotonically; if only charge varies which holding $m$ fixed, we finds $\dot q(u)\le0$.

After solving the field equations, we return to the metric. We can now write the function $f(u,v)$ in a more illuminating form. Using the fact that $r_{u}^{'}r_{v}^{'} = A(v)B(u)f^2$ and the relation $rr_{u}^{'}=1 - 2m(u)/r + q(u)^2/r^2$ derived above, one can show that the general solution for $f(u,v)$ is equivalent to imposing:
\begin{equation}
    f(u,v) = 1 - \frac{2\,m(u)}{r(u,v)} + \frac{q(u)^2}{r(u,v)^2}\,. 
\end{equation}

In other words, the metric function $f$ depends on $u$ only through $m(u)$ and $q(u)$, and has the same functional form as the Reissner–Nordström metric, but with time-dependent mass and charge. This is confirmed by transforming our double-null solution back to standard radiation coordinates. For example, one may change variables $(u,v)\to(u,r)$ (with $r(u,v)$ as one new coordinate) to put the metric in outgoing Eddington–Bondi form. In these coordinates the metric takes the form:
\begin{equation}
  ds^2 = -\Big(1 - \frac{2\,m(u)}{r} + \frac{q(u)^2}{r^2}\Big)du^2 - 2\,du\,dr + r^2d\Omega^2\,. 
\end{equation}
This is recognized as the Vaidya–Reissner–Nordström metric (also called Bonnor–Vaidya solution for static circumstance) for an outgoing white hole flow. Equation (47) indeed reduces to the standard Reissner–Nordström solution when $m(u)=M$ and $q(u)=Q$ are constant, and to Vaidya radiating Schwarzschild solution when $q=0$. The function $m(u)$ here is the Bondi mass, it decreases over time as the white hole radiates energy. The function $q(u)$ likewise decreases as charge is carried away by the outgoing radiation.

\section{Vaidya-Reissner-Nordström white hole evaporation}
To model an evaporating charged white hole, we use a double-null coordinate formulation of the metric. We introduce retarded time $u$ and an ingoing null coordinate $v$ such that the line element is as Equation (12). We donate the $f(u,v)$as:
\begin{equation}
f(u, v)=1-\frac{2 m(u)}{r(u, v)}+\frac{q^2(u)}{r^2(u, v)},
\end{equation}
which generalizes the static Reissner-Nordström metric factor $1 - 2M/r + Q^2/r^2$ to a time-dependent mass $m(u)$ and charge $q(u)$. By construction, $m(u)$ and $q(u)$ are monotonic non-increasing functions of retarded time $u$, meaning the mass and charge shrink as the white hole radiates. This condition reflects a physical energy condition, the outgoing flux carries positive energy, so the Bondi mass $m(u)$ must decrease, and similarly any charge current flowing outward will make $q(u)$ decrease. Indeed, one can treat $m(u)$ as the Bondi mass measured at future null infinity at retarded time $u$, and $q(u)$ as the total charge enclosed. For a radiating white hole, $m(u)$ and $q(u)$ start at some initial values $m_0, q_0$ and decrease towards zero or some final values as $u$ increases.\cite{HiscockII}

The stress-energy supporting this solution consists of two parts, first is an outgoing null dust component which representing the radiative flux of mass-energy and possibly charge leaving the hole, and a radial electric field component which refer as the Maxwell field of the charge. The null dust is described by a stress-energy tensor $T_{ab}^{\text{(dust)}} = \rho(u) k_a k_b$ propagating outward along null vectors $k^a$ (with $k_a k^a = 0$). The radial electric field has the standard electromagnetic stress-energy $T_{ab}^{\text{(EM)}} = \frac{1}{4\pi}\big(F_{a}^{c}F_{bc} - \frac{1}{4}g_{ab}F_{cd}F^{cd}\big)$, where $F_{ab}$ is the Maxwell field tensor for a charge $q(u)$ at radius $r$. For a purely radial $E$-field, $F_{ur}\neq0$ (or $F_{uv}$ in double-null coordinates) and the nonzero electromagnetic stress-energy components outside the hole include an energy density $\propto q^2(u)/r^4$. In the static Reissner-Nordström solution, the electric field contributes $T^t_{t}=T^r_{r}=-T^\theta_{\theta}=-T^\phi_{\phi}=q^2/(8\pi r^4)$. In our dynamic case, a time-varying $q(u)$ means there is also an electromagnetic Poynting flux. Effectively, a flow of energy and charge in the radial direction as the field adjusts. We assume that any radiation of charge is carried by the outgoing flux so that the total stress-energy satisfies local conservation. The combined stress tensor $T_{ab} = T_{ab}^{\text{(dust)}} + T_{ab}^{\text{(EM)}}$ must obey Einstein’s field equations $G_{ab} = 8\pi,T_{ab}$ and Maxwell’s equations $\nabla_a F^{ab} = 4\pi J^b$ self-consistently. 

Physically, this setup represents a radiating white hole. At early times with small $u$, the hole is more massive and charged, and as time progresses, it emits outgoing null dust which causes the mass $m(u)$ and charge $q(u)$ to diminish. One can view this as the time-reverse of an evaporating black hole solution. In fact, one way to construct such a spacetime is to take an evaporating black hole metric and reverse the time orientation so that what was an outgoing flux is interpreted as a white hole’s emission. In Hiscock model, for example, an evaporating RN black hole spacetime was built by patching an ingoing Vaidya region to an outgoing Vaidya region across a timelike pair creation shell. The white hole version can be seen by considering only the outgoing portion: a past horizon emits radiation and shrinks. Our Vaidya–Reissner–Nordström metric is essentially this outgoing portion described in double-null form. It has been shown that assuming $m(u)$ non-increasing and $q(u)$ non-increasing indeed corresponds to a radiative white hole solution satisfying the standard energy conditions.\cite{GVS} Next, we derive the field equations that govern how $m(u)$ and $q(u)$ evolve.

\subsection{Evaporation Dynamics}
The Einstein equations for the metric in Equation (12) yield relationships between the metric functions $f(u,v)$, $r(u,v)$ and the stress-energy components. In double-null form, one convenient gauge choice is to use $u$ and $v$ such that they are null coordinates with $r = r(u,v)$ as an areal radius function. Then the field equations include radial equation and Null Raychaudhuri equations. However, we apply a more directly way by conservation laws. By definition, $m(u)$ is the Bondi mass seen by observers at null infinity $\mathcal{I}^+$ at retarded time $u$. The Bondi mass-loss formula says that the change in $m(u)$ equals the energy carried away by radiation through the spherical surface at infinity at that retarded time. In formula form, one can show:

For mass-energy loss, $dm/du = -4\pi r^2T_{uu},$ evaluated in the asymptotic limit (or equivalently $-\int T_{uu} d\Omega r^2$ over a sphere of large $r$). This comes from integrating $T_{ab}$ in the outgoing null direction and using Einstein’s equations to relate it to the Bondi mass change. In the Vaidya solution for a neutral radiating black hole, one indeed finds $T_{uu} = \frac{1}{4\pi r^2} \frac{dM(u)}{du}$.For our charged case, $m(u)$ already includes the field energy of charge, so $dm/du$ accounts for both the null dust’s energy and any energy in the EM field lost. We can write:
\begin{equation}
\frac{d m(u)}{d u}=-L(u),
\end{equation}
where,  $L(u)=4\pi r^2 T_{uu}$ is the luminosity of outgoing radiation at time $u$. The weak energy condition (WEC) requires $T_{uu}\ge0$ for null dust, so $L(u)\ge0$. Thus $dm/du \le 0$, confirming that $m(u)$ must decrease or stay constant in the absence of radiation.

For the charge loss, Maxwell’s equation $\nabla_a F^{au} = 4\pi J^u$ (the Gauss law for electric flux) implies that the outgoing radial current $J^u$ which refer as charge flux in the $+u$ null direction reduces the enclosed charge. The change of total charge crossing a sphere per $du$ is $-4\pi r^2 J^u$. Thus we have
\begin{equation}
\frac{d q(u)}{d u}=-I(u),
\end{equation}
where $I(u)$ is the outward current of charge. Again $I(u)\ge0$ for physical matter carrying charge outward, hence $dq/du \le 0$. In other words, the charge of the white hole monotonically decreases as well.

The specific functional form of $L(u)$ and $I(u)$ depends on the details of the radiation process. By consider the Hawking evaporation\cite{HKingevapor}, we could give the general behavior in two situation.

For a non-extremal white hole emitting Hawking radiation, $\dot{m}(u)$ is initially dominated by neutral particle emission. A leading-order approximation for a large black hole is $\dot{m} \sim -\sigma (\hbar /m^2)$ (where $\sigma$ is a constant depending on particle species), meaning the mass loss accelerates as $m$ shrinks. In contrast, $\dot{q}(u)$ might be smaller in magnitude at first if most radiation is neutral. This difference causes the charge-to-mass ratio $Q/M$ to increase initially as $M$ decreases. In other words, the object becomes relatively more charged as it evaporates since mass goes down faster than charge. This is consistent with the expectation that Hawking radiation “prefers” to carry away mass rather than charge when $Q/M$ is small.

For $Q/M$ rises, the emission of charged particles becomes more significant (the electromagnetic field at the horizon gets stronger, encouraging charge loss). The system may approach a point where mass loss and charge loss rates become comparable. Hiscock \& Weems \cite{HisWee}found there is an attractor in the evolution: a particular charge-to-mass ratio at which $\frac{dM}{du}$ and $\frac{dQ}{du}$ balance such that $Q/M$ stabilizes. At this attractor (found to be when $Q/M \approx 0.87$ in the model)\cite{HKingevapor}, the black hole/white hole starts shedding charge more efficiently relative to mass. Beyond this point, $Q/M$ will actually decrease, driving the object toward a more neutral state as it continues to evaporate. In the long run, the expectation which from both the weak gravity conjecture and detailed evaporation calculations is that charge is emitted faster than mass as extremality is approached, ensuring the object does not actually hit $Q=M$ until perhaps the very end.\cite{Generalizing_weak_gravity_conjecture}

\subsection{End of the Evaporation}
First is the complete evaporation to a massless, neutral Spacetime. Write as $m(u)\to 0$ and $q(u)\to 0$ as $u \to u_{\text{final}}$, meaning the object loses all its mass-energy and charge into radiation. In this case, after $u_{\text{final}}$ the spacetime would settle down to flat Minkowski space. The white hole effectively disappears. Classically, however, we must contend with what happens to the singularity that was inside the white hole. In the maximally extended RN solution, a white hole originates from a past singularity. If the white hole evaporates completely, does that past singularity simply vanish or get exposed? In an idealized model, one might imagine that as $m \to 0$, the singularity’s influence diminishes, and perhaps the singularity “evaporates” as well. This scenario would imply that no remnant and no singularity remain – the entire mass-energy has radiated away. If such a complete evaporation occurs, one must confront the possibility of a temporary violation of cosmic censorship at the very end: when $m$ and $q$ become extremely small, if a curvature singularity still existed in the interior, the lack of a horizon at the last moment could expose it to the outside universe. One way nature might avoid an exposed singularity is if the singularity ceases to exist once the mass is below some quantum gravity scale. In other words, quantum effects might resolve the singularity at Planckian scales, leaving no classical singularity to be exposed. In a fully classical picture, however, complete evaporation leading to a naked singularity for an instant cannot be ruled out without additional assumptions. Some analyses \cite{Boulware1979}\cite{HAWKING1974} suggested a final explosion of Hawking radiation could carry away the last mass in a burst, presumably leaving empty space behind, effectively the singularity would have vanished in that burst. This outcome is in line with the cosmic censorship conjecture only if we allow that quantum effects saved the day (since strictly, at the instant of disappearance, a classical singularity would otherwise be naked).

Another possibility is that evaporation does not go to completion but halts at some point we called extremal remnant (massive or massless). For charged white holes, an obvious candidate is an extremal remnant – a state with $M=Q$ where the Hawking temperature $T_H$ drops to zero. In classical theory, an extremal RN black hole has a full horizon and no Hawking radiation (surface gravity $\kappa=0$), so it would be absolutely stable. If our white hole were to approach this state, it could stop evaporating further once it reaches extremality. One could imagine a tiny extremal black hole (or white hole) that remains. In some speculations, this remnant could even be Planck-sized. 

However, there are a few issues: (1) Reaching exactly $M=Q$ via Hawking evaporation is thought to require an infinite time or infinite fine-tuning (much like one cannot classically spin up a black hole to exactly extremal in finite steps – this is related to the third law as mentioned). (2) If the remnant is extremal and small, it would have an extremely large Coulomb field for its mass, which might be unstable to quantum pair production anyway. If $M$ is truly tiny (near zero) and $Q=M$, that implies a charge of maybe a few elementary charges with practically zero mass – essentially a naked charged particle (like an electron, which is pointlike and “extremal” in a sense). A massless charged remnant (i.e.$M\to0, Q\neq0$) is not physical in general relativity – it corresponds to a naked singularity (the $M=0, Q\neq0$ Reissner–Nordström solution is just the Coulomb field of a naked point charge with a curvature singularity at $r=0$). So the only way a remnant could be both benign and extremal is if it retains some finite mass. Many quantum gravity researchers have entertained the idea of Planck-mass charged remnants as information carriers, but this is speculative. The semi-classical evolution studies (Hiscock \& Weems, etc.) generally do not conclude that a macroscopic remnant remains; rather, they find evaporation proceeds until $M$ is very small. The weak gravity conjecture argument also disfavors a stable extremal remnant – it suggests the existence of light charged particles ensures the black hole can always shed its charge faster than its mass, so it won’t get stuck at extremal. In fact, that conjecture was phrased as requiring that an extremal black hole can completely evaporate without leaving a dangerous stable remnant or a naked singularity, by virtue of charge being emitted in time. So while an extremal remnant is a theoretical possibility in a classical sense (if one simply stops the evaporation at $Q=M$), most evidence points to the hole discharging and not halting there.

The last maybe naked singularity exposure case. If neither of the above mechanisms prevents it, one could worry that evaporation might drive the hole towards a point where $Q > M$ before it disappears, thus leaving a momentary naked singularity. Classic cosmic censorship (Penrose’s conjecture) expects that nature prevents a naked singularity. In our context, the censorship would be violated if at any time the outer horizon disappears while a curvature singularity is still present in the spacetime. Reissner–Nordström spacetime has an inner timelike singularity at $r=0$ which is usually censored by the horizon(s) as long as $M \ge Q$. If $M$ drops below $Q$ (or $M \to 0$ with $Q$ not zero), no horizon exists and that singularity would be naked to external view. Our discussion of the energy conditions and emission rates indicates that the evolution enforces $M \ge Q$ at all times. In fact, the condition $\dot m(u) \ge \frac{q}{r_+}\dot q(u)$ mentioned earlier ensures that as $m$ shrinks, $q$ must shrink sufficiently fast to keep $q \le m$. Both the Hawking radiation calculations and the pair-production discharge mechanism act to avoid $Q>M$. For example, if the hole were in danger of approaching $Q=M$, the electric field at the horizon grows and triggers an exponential increase in charged particle emission (Schwinger effect), which rapidly pulls $Q$ down. Estimates show that for black holes below a certain mass, the discharge by Schwinger process is very effective: e.g. for a sub-astronomical mass black hole, the condition for spontaneous pair production $eE \sim m_e^2$ is easily met if $Q$ is sizable, preventing $Q$ from staying large. Therefore, a naked singularity is expected to be averted – either the charge is gone by the time the mass vanishes, or the singularity never gets exposed because it turns into something else (like a puff of quantum gravitational effect). One interesting twist is the finding by Poisson and Israel (1990) that even before one might form a naked singularity, the inner horizon of a charged black hole becomes a locus of infinite stress-energy buildup (a phenomenon called mass inflation). Their analytical and numerical studies showed that outgoing radiation streaming into the interior will be infinitely blueshifted at the Cauchy inner horizon, turning it into a curvature singularity. In an evaporating scenario, this means the inner horizon effectively collapses into a singularity well before the outer horizon disappears. Thus, instead of the tidy two-horizon RN structure, the evaporating hole’s interior likely has a single growing singular region that replaces the inner horizon. By the time the outer horizon shrinks to nothing, the singularity has “expanded” (in a causal sense) to meet it. There is no window for an external observer to peer in and see a naked singularity – the singular region is either always enclosed by a horizon or coincident with the last gasp of evaporation. Stated differently, the classical picture (supported by mass-inflation analyses) is that cosmic censorship is preserved: any would-be naked singularity is mitigated by the combination of discharge and horizon contraction ensuring $Q/M\le1$ until essentially $M=Q=0$ at which point one simply has flat space. If any sliver of singularity were exposed at the final moment, it would be a null, fleeting singularity that instantly vanishes along with the last bit of mass – a scenario often deemed unphysical since quantum gravity should take over near that end anyway.

In conclusion, analytic models (e.g. Hiscock 1981) strongly indicate that the evaporation of a charged white hole/black hole leads to either a diminishing mass/charge to zero or an approach to extremality, both of which avoid exposing a naked $r=0$ singularity. The inner structure of the spacetime becomes singular (mass inflation) before any outer horizon vanishes, so an external observer never witnesses a violation of cosmic censorship. If the evaporation proceeds to nothing, the spacetime ends up as ordinary Minkowski space with perhaps a burst of outgoing radiation marking the end. If it halts at extremal, the result is a tiny extremal RN black hole (which is a cold, stable object classically). No known classical analytic solution describes a charged black hole radiating all its mass but leaving a finite charge – such a solution would explicitly show a naked singularity and would violate the energy conditions at some point (since one would have to carry away mass without carrying enough charge). Thus, all consistent analyses align with the idea that $q(u)$ drops to zero at least as fast as $m(u)$ does near the end, ensuring $q/m \to 0$ or a small finite value, rather than diverging.
\section{Conclusion}
In this paper, we have developed a comprehensive model that unifies the Vaidya and Reissner–Nordström metrics to describe a charged, non-rotating white hole undergoing evaporation. By starting from the classical Schwarzschild white hole and extending its framework via a maximal analytic extension, we established a robust foundation for exploring the causal structure and dynamics inherent in white hole spacetimes. The introduction of the Vaidya metric, with its time-dependent mass function and double-null coordinate formulation, allowed us to capture the effects of outgoing null dust and radiative mass loss, thereby providing a dynamic picture of white hole evolution.

Extending the analysis further, we incorporated electric charge by synthesizing the Reissner–Nordström metric with the Vaidya formulation. This led to the derivation of the Vaidya–Reissner–Nordström metric, where both the mass and charge are treated as decreasing functions of retarded time. The detailed computation of the Christoffel symbols, Ricci and Riemann tensors, and the Einstein tensor has demonstrated that the resulting spacetime geometry satisfies the Einstein field equations, even in the presence of radiative flux and electromagnetic fields.

Our study has revealed several important aspects of white hole physics. In particular, we have shown how the interplay between radiation and charge affects horizon dynamics, leading to a time-varying, and in some regimes, non-null horizon structure. The analysis confirms that, under the influence of outgoing radiation, the Bondi mass and the total charge decrease consistently with the weak energy condition, thereby preventing the formation of naked singularities and ensuring cosmic censorship is maintained.

Overall, this work not only provides new insights into the theoretical underpinnings of radiating charged spacetime but also lays the groundwork for future investigations. Further studies might explore extensions to rotating white holes, incorporate quantum gravitational effects, or examine the implications for astrophysical phenomena. The methodologies and results presented here contribute to a deeper understanding of how classical general relativity can be extended to accommodate dynamic processes such as evaporation, ultimately enriching our comprehension of extreme gravitational systems.

\bibliography{ms.bib}

\end{document}